\documentclass[preprint,review ,authoryear,11pt]{elsarticle}

\usepackage{amssymb,multirow,amsmath,graphicx,array,adjustbox,booktabs,hyperref,rotating}
\usepackage[no-weekday]{eukdate}
\usepackage{geometry}
\usepackage[usenames,dvipsnames]{color}

\usepackage[normalem]{ulem}

\usepackage{mathpazo,parskip}
\begin{document}

\begin{frontmatter}

\title{Forecasting for Social Good}

  \author[label1]{Bahman Rostami-Tabar \corref{cor1}}
 \ead{rostami-tabarb@cardiff.ac.uk}
 \cortext[cor1]{Correspondance: B. Rostami-Tabar, Cardiff Business School, Cardiff University, Cardiff, CF24 4YX, UK. Tel.: +44-(0)29 2087 0723}
 \address[label1]{Cardiff Business School, Cardiff University, Cardiff, CF24 4YX, UK}
 \author[label2]{Mohammad M Ali}
 \address[label2]{Royal Docks School of Business and Law, University of East London, London, E16 2RD, UK}
 \author[label3]{Tao Hong}
 \address[label3]{ Energy Production and Infrastructure Center, Univ. of North Carolina Charlotte, Charlotte, NC 28223, USA}
  \author[label4]{Rob J Hyndman}
 \address[label4]{Department of Econometrics \& Business Statistics, Monash University, Clayton VIC 3800, Australia}
  \author[label5]{Michael D Porter}
 \address[label5]{Department of Engineering Systems and Environment, University of Virginia, Charlottesville, VA 22904, USA}
    \author[label1]{Aris Syntetos}

\begin{abstract}
Forecasting plays a critical role in the development of organisational business strategies. Despite a considerable body of research in the area of forecasting, the focus has largely been on the financial and economic outcomes of the forecasting process as opposed to societal benefits. Our motivation in this study is to promote the latter, with a view to using the forecasting process to advance social and environmental objectives such as equality, social justice and sustainability. We refer to such forecasting practices as Forecasting for Social Good (FSG) where the benefits to society and the environment take precedence over economic and financial outcomes. We conceptualise FSG and discuss its scope and boundaries in the context of the ``Doughnut theory''. We present some key attributes that qualify a forecasting process as FSG\@: it is concerned with a real problem, it is focused on advancing social and environmental goals and prioritises these over conventional measures of economic success, and it has a broad societal impact. We also position FSG in the wider literature on forecasting and social good practices. We propose an FSG maturity framework as the means to engage academics and practitioners with research in this area. Finally, we highlight that FSG: (i)~cannot be distilled to a prescriptive set of guidelines, (ii)~is scalable, and (iii)~has the potential to make significant contributions to advancing social objectives. 

\end{abstract}


\end{frontmatter}

\section{Background and motivation}

Organisations make operational, tactical and strategic decisions every day. Regardless of the sector or industry, these decisions reflect the expectations of what the future may look like. This is where forecasting can play a crucial role as an integral part of a decision-making process \citep{hyndman2018forecasting}. This is well understood in areas with commercial or economic interests. Forecasting and its link to business decision-making has been under research for decades \citep{gonzalez2016forecasting,sanders2016forecasting,gilliland2016business,ord2017principles}. Many important contributions have been offered in these fields (e.g., macroeconomics and the financial sector, retail industry and supply chains, energy industry and tourism \citep{fildes2002state,fildes2008forecasting,syntetos2009forecasting,athanasopoulos2011tourism,hong2014global}) on how forecasting may improve organisational decision-making. However, such studies have largely sought to improve forecasting processes (and their integration with decision-making) in the presence of financial or economic motivations. On the other hand, little attention has been paid to forecasting when the emphasis is on deriving some societal benefits regardless of the financial or economic implications. In this article, we refer to such forecasting practices as \emph{Forecasting for Social Good} (FSG).

While there is a growing recognition by agencies, organisations, and governments that data-driven decision-making tools, such as forecasting models, may offer significant improvements to society \citep{iyer2014reshaping}, there is not a cohesive body of research that offers guidance towards the conceptualisation, implementation and evaluation of forecasting models for social good in practice. Although some work has been done in this area \citep{gorr2003introduction,nsoesie2014systematic,van2016demand,wicke2019using,litsiou2019relative}, progress has been relatively slow and sporadic, both in terms of academic contributions and practical applications. This is exemplified by the fact that the development and use of forecasting models in organisations with social missions (especially in health, humanitarian operations and the third sector) is considerably under-developed. Evidence \citep{getzen2016measuring,cacciolatti2017clashing,lu2018empirical} suggests that this may be due to a lack of awareness, skills and understanding of the value of forecasting, but the fact remains that such organisations are largely not exploiting  (relevant) forecasting capabilities.
Further, major review papers in the areas of forecasting, as well as operations research and operations management when forecasting is explicitly considered \citep{fildes2008forecasting,syntetos2009forecasting,boylan2010spare,syntetos2016supply,makridakis2020forecasting}, do not take into account work related to FSG\@. The paucity of academic contributions may be due to the limited amount of existing work to build upon, or the fact that relevant work might appear in journals not frequently read by the forecasting community \citep{soyiri2013overview,nsoesie2014systematic,dietze2017ecological,goltsos2019forecasting}.
Given the background discussed above, we feel it is timely to explicitly address the definition of FSG and its positioning in the wider body of knowledge. This exercise will facilitate the discussion of both forecast implementation and evaluation issues leading to the proposition of a research agenda; it should also allow organisations to advance their social missions and benefit from the value forecasting may offer. The purpose of this paper is three-fold:
\begin{itemize}
    \item increase awareness and interest of academics and practitioners on the potential impact of FSG;
    \item encourage interested academics and practitioners to engage in the FSG agenda;
    \item inspire the development of new forecasting methodologies tailored for social good applications.
\end{itemize}

The remainder of the article is organised as follows. Section~\ref{sec:fsg} defines the area of FSG, its scope and boundaries as well as its relation to (other) data-driven social good initiatives and forecasting areas. Section~\ref{sec:quad} suggests a positioning framework on the basis of (i) the maturity of the forecasting process (theory) and (ii) the use of forecasting in social good (practice). It also provides an indicative agenda for further research. Finally, Section~\ref{sec:concl} presents a summary of our conclusions.

\section{Forecasting for Social Good}\label{sec:fsg}

In this section we first explain the Doughnut theory used to frame our definition and scope of FSG\@. This theory is an alternative way of looking at growth economies. It prioritises people and the planet over economic growth, which can help us as a society thrive within the limits of our planetary boundaries \citep{raworth2017doughnut}. In this paper, the theory helps to create a common understanding of the term Forecasting for Social Good. 

We attempt to answer the following two questions:
\begin{enumerate}
    \item What is meant by FSG?
    \item What attributes/features make a forecasting process aligned with FSG? That is, when does a forecasting process belong to FSG and when does it not?
\end{enumerate}

\subsection{Doughnut theory}\label{sec:Doughnut}

\begin{figure}[!htb]
\centerline{
\includegraphics[width=.6\linewidth]{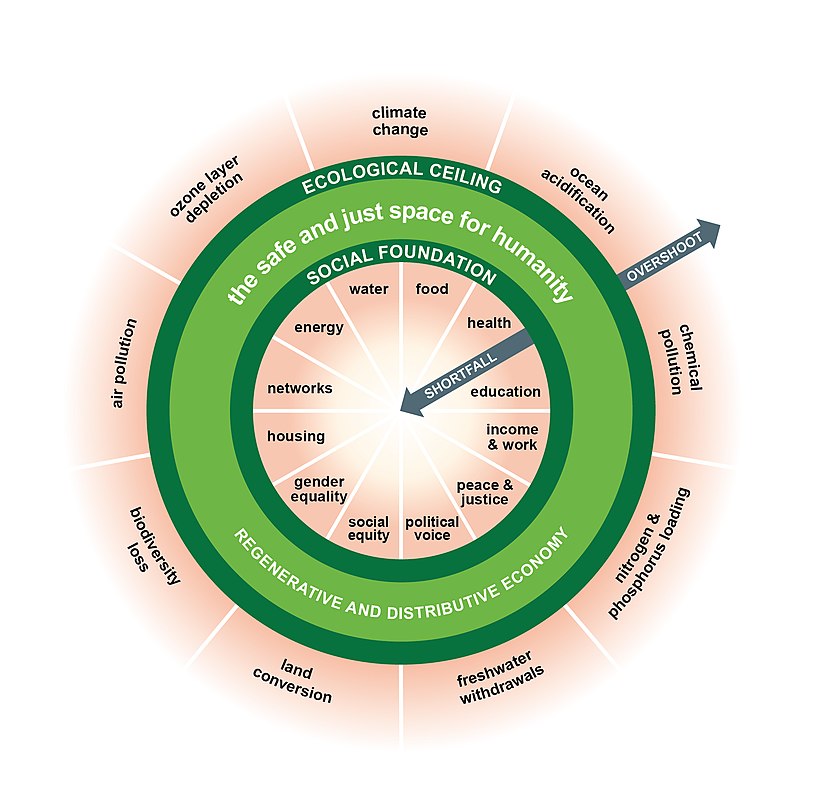}}
\caption{The classic image of Doughnut with social and planetary boundaries, Source: \citet{wiki:xxx}.}
\label{fig_doughnut}
\end{figure}

Doughnut theory was proposed by \cite{raworth2017doughnut} and offers a framework for thinking about how we create a world in which humanity thrives. Raworth states that, “instead of economies that need to grow, whether or not they make us thrive, we need economies that make us thrive, whether or not they grow”. The aim is to meet the needs of all people within the means of the living planet. The theory combines the concept of social foundation with that of ecological ceiling in a single framework as illustrated in Figure~\ref{fig_doughnut}.

The social foundation is derived from the social priorities described in the United Nations Sustainable Development Goals \citep{un2015goals}. The idea is to ensure that no one is left in the hole of the doughnut below the social foundation and falls short on essentials of life ranging from food and clean water to gender equality, and everyone has a political voice and access to housing.

The ecological ceiling includes nine planetary boundaries developed by environmental scientists \citep{rockstrom2009safe} that represent the planet’s capacity of critical life-supporting systems. In order to preserve them, humanity must live within these ecological boundaries while meeting the needs of all described in the social foundation.

Between the social foundation and the ecological ceiling lies a an space in which it is possible to meet the needs of all people within the means of the living planet --- an ecologically safe and socially just space in which humanity can thrive.

This is the space we must move into from both sides simultaneously, in ways that promote the well-being of all people and the health of the whole planet. Achieving this globally calls for action on many levels, including research and its applications. The framework has been adopted in multiple academic disciplines, various countries, sub-regions and cities worldwide \citep{cole2014tracking,dearing2014safe,hoornweg2016urban,amenta2020experimenting,bennett2020reorienting}.

\subsection{Definition and scope of Forecasting for Social Good}\label{sec:scope}

The Doughnut framework allows multi-metric ‘compasses’ to be elaborated for informing the decision-making process \citep{dearing2014safe}. In order to promote the well-being of all people and the health of the whole planet, the decision-making process needs to support all activities that bring us into the Doughnut space --- an environmentally safe and socially just space --- in which humanity thrives. We note that one of the main components of any decision-making process is forecasting. 

We define forecasting as a genuine prediction of the future, given all the information available at the time the forecast is generated, including historical data and knowledge of any future events that might impact the outcome(s) \citep{goodwin2018profit,hyndman2018forecasting}. The forecasting process starts by taking inputs in the form of a problem description, data and information, then an appropriate forecasting method is identified and the inputs are processed and formulated to implement the method using a software and make the forecast, incorporating human judgement and uncertainty assessments when necessary. 

\begin{figure}[!htb]
\centerline{
\includegraphics[width=.5\linewidth]{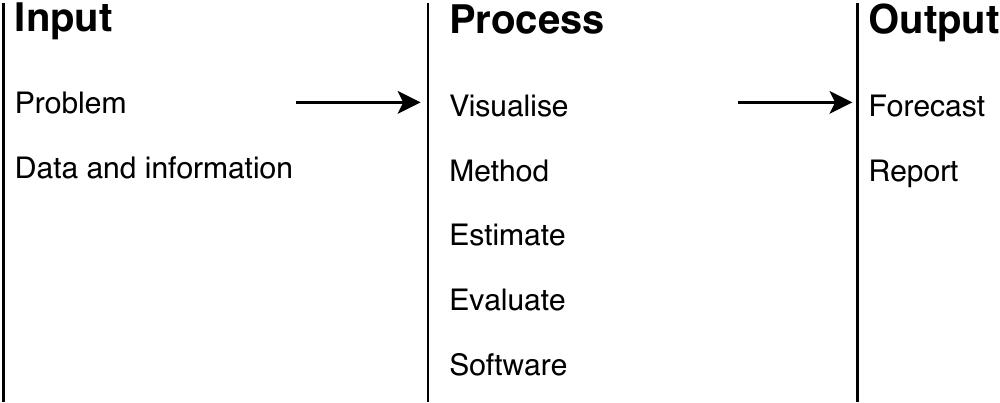}}
\caption{Forecasting Process}
\label{fig_forecasting_process}
\end{figure}

Genuine forecasting can also take place in the absence of available data and not relying on statistical methods or using statistical software. Instead, we may rely on structured management judgement including the Delphi method, forecasting by analogy, surveys, scenario forecasting and other judgemental forecasting approaches.

\begin{figure}[!htb]
\centerline{
\includegraphics[width=.7\linewidth]{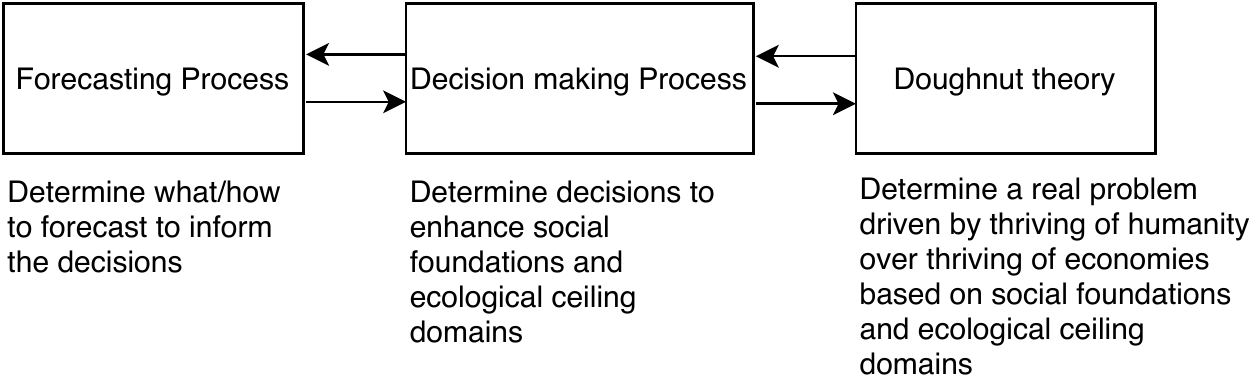}}
\caption{Forecasting for Social Good Process}
\label{fig_process}
\end{figure}

Forecasting is used to help decision makers to make more informed and potentially better decisions. Therefore, forecasts need to be tailored to provide answers to the questions a decision maker needs in a particular set of circumstances. In the case of FSG, we argue that the forecasting process should be determined by a decision-making process that leads a community into an ecologically safe and socially just space where it can thrive. Figure~\ref{fig_process} shows the relationship between the Doughnut theory, decision-making process and the forecasting process in FSG.

FSG is a forecasting process that aims to inform decisions that prioritise thriving of humanity over thriving of economies by enhancing the social foundation and ecological ceilings that impact public as a whole at both local and global levels. Therefore, FSG contributes to the solutions to real problems that are primarily driven to thrive humanity by enhancing the social foundation within the planetary capacity. While profits and other growth-oriented metrics can be considered they are not given priority.

Now we move towards our second question i.e. what attributes make a forecasting process a FSG. We argue that to qualify for FSG, a forecasting process needs to have four attributes: (i) it is concerned with a real problem; (ii) the problem is primarily driven by thriving humanity over thriving of economies; (iii) the proposed solution enhances the social foundation and ecological ceiling; and (iv) it impacts the public as a whole. These are further discussed below.

\textbf{Real Problem}: 
FSG emphasises the problems directly affecting people/humanity and are experienced in daily life, in contrast to the problems mostly residing in the theoretical world. While the scope of other similar initiatives such as Data Science for Social Good \citep{inproceedings}  might be limited to real problems in sectors such as government and/or the voluntary sector, our definition of FSG is inclusive and encompasses all organisations irrespective of the industry and whether they are governmental, commercial or voluntary organisations. Hence, the scope and the nature of the problems that the forecasting process is attempting to provide solutions for could range from a task in a profit-driven organisation e.g. forecasting to reduce waste, to a whole sector, e.g. forecasting for humanitarian and disaster relief operations. This is important as commercial organisations are rapidly changing in terms of how they think and position themselves when it comes to social good, and they should not be excluded in the definition \citep{rostami2019isf}.
This dimension highlights an important aspect of FSG - that is the collaborative effort and continuous interaction between the problem owner and the forecaster to define the problem, design the model, evaluate and implement the solution and link it to the decision-making process. The collaborative efforts will lead to questions that are not only crucial to help humanity to thrive but also provide opportunities for innovative research.

\textbf{Prioritise thriving of humanity over thriving of economies}: The second attribute focuses on the objectives of solving the real problems under consideration. FSG’s outputs prioritise thriving of humanity over the thriving of economies. Therefore, one of the key features that define FSG is whether the purpose of informing decisions -by the forecasting process- to solve the real problem, is driven primarily by social/environmental considerations or economic growth.
FSG is not primarily driven by economic growth i.e. the goal is to help humanity thrive within environmental boundaries whether the economy grows or not. The is a radical change in the way we look at forecasting process. The idea is to ensure that decisions and actions informed by forecasts are helping humanity to get into the doughnut-shaped space, an ecologically safe and socially just space for humanity to thrive in. The forecasting process may also result in economic growth. However, it is within the scope of FSG if the primary focus is to improve the human and planetary condition.

\textbf{Enhance social foundation within ecological ceiling}: The third dimension of FSG relates to how the benefits of the forecasting outputs are being measured. In a traditional business forecasting scenario, the outputs or the empirical utility will be associated with the financial or economic implications. However, in the case of FSG, the forecasting process focuses on the social foundation as the primary output. Forecasting should inform decisions towards enhancing social foundation while maintaining or improving the ecological ceiling simultaneously. Therefore, we need indicators and metrics that allow us to measure both components.
Doughnut’s social foundation includes twelve dimensions that are derived from internationally agreed minimum social standards described in the Sustainable Development Goals (SDG) defined by the United Nations \citep{un2019report}. SDG indicators are relatively well thought through at an international level and developed/refined by hundreds of multidisciplinary experts. Also, they are already being integrated into national and transnational policies as well as referenced in academia \citep{cancedda2018time,biermann2017global}. Doughnut’s social foundation  include water, food, health, education, income \& work, peace and justice, political voice, social equity, gender equality, housing, networks and energy. Various metrics such as nutrition, sanitation, income, access to energy, education, social support, equality, democratic quality, employment, self-reported life satisfaction and healthy life have been used in various studies to quantify social foundation \citep{steinberger2010constraint,cole2014tracking, dearing2014safe, raworth2017doughnut,o2018good}.

The ecological ceiling consists of nine dimensions that are vital to our planet’s ability to sustain human life as set out by \cite{rockstrom2009safe}. Beyond these boundaries lie unacceptable environmental degradation and potential tipping points in Earth systems. These boundaries include ozone layer depletion, ocean acidification, nitrogen and phosphorus loading, chemical pollution, freshwater depletion, land conversion, air pollution, climate change and biodiversity loss. Indicators used in various studies include phosphorus, nitrogen, ecological footprint, material footprint, CO2 emissions, greenhouse gas emissions \citep{knight2011environmental,dearing2014safe, lamb2015human, o2018good}.

When a forecast is made to inform a decision, the penalty will arise if the forecast turns to be different from the actual value. The idea in FSG is to use amended penalty functions that integrate social foundation and ecological ceiling indicators instead of current functions based on statistical, economical and financial KPIs \citep{berk2011asymmetric,lee2008loss}. FSG informs decisions that enhance social foundation indicators and not violate any principle measures of ecological ceiling. There is still more to be done to define new metrics for social foundation and ecological ceiling at local and global levels and this is one of the important challenges facing humanity.

Traditionally, forecasting publications, conferences and practices focus on methodological advances and profit driven goals. This would need a radical shift to allow researchers and practitioners to get involved in FSG research.

\textbf{Impact on the public as a whole}: The last dimension focuses on who benefits from the application of forecasting. FSG gives priority to both local and global levels rather than focusing only on its local beneficiaries themselves. FSG can be used at multiple scales – from an individual to a nation –-- as a tool for transformative action that embraces social and ecological metrics, both locally and globally. Organisations should ensure that these metrics are measured through the internal activities rather than external activities such as donation to a charity.

\begin{figure}[!htb]
\centerline{
\includegraphics[width=.6\linewidth]{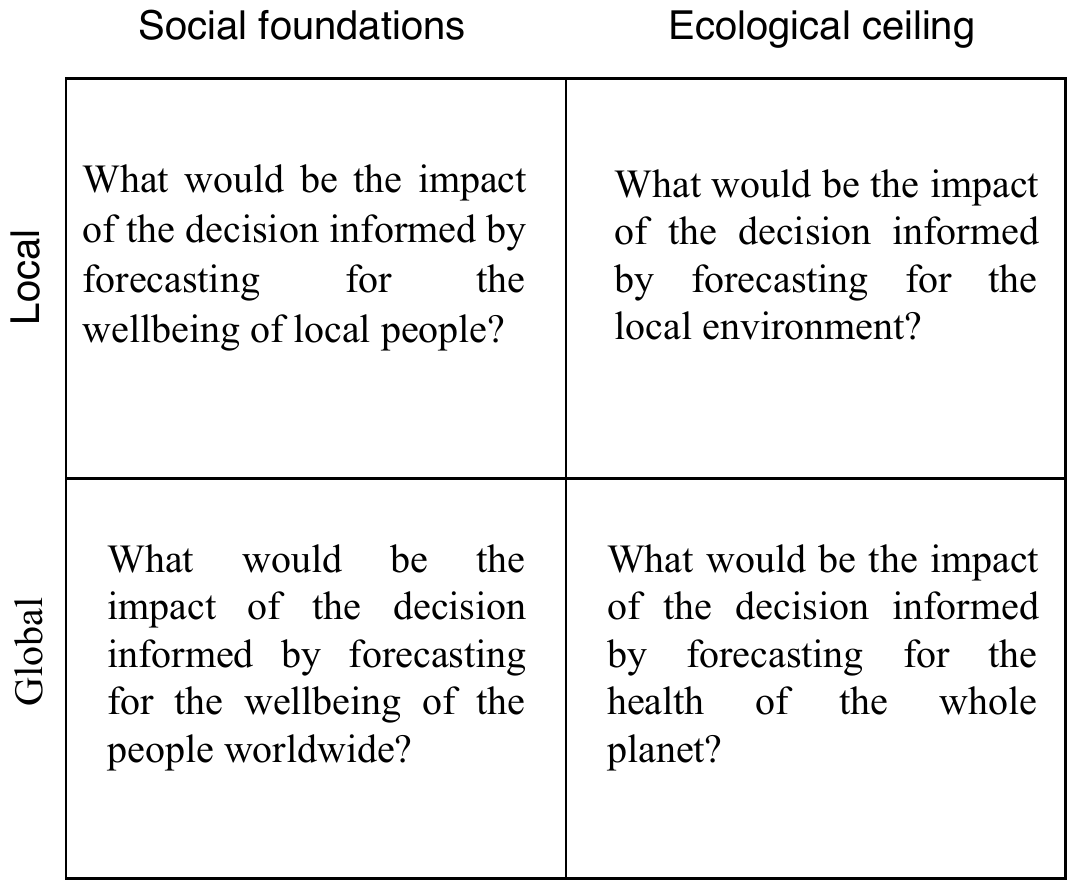}}
\caption{FSG beneficiaries.}
\label{fig_public}
\end{figure}

FSG starts by asking this question: How can the forecasting process inform decisions that help thriving humanity whilst respecting the wellbeing of all people, and the health of the whole planet? Following this question, the benefit of FSG can be assessed across four lenses that arise from combining two type of benefits (social foundation and ecological ceilling) and two scales (local and global) as depicted in Figure \ref{fig_public}.

In this section, we first clarified what is meant by Forecasting for Social Good (FSG) and then moved towards defining the four attributes of FSG. Any forecasting process can qualify as FSG if it focuses on a real problem, is primarily driven by thriving of humanity over thriving of economies, it enhances social foundation and ecological ceiling, and impacts the public as a whole at both local and/or global levels.

\begin{figure}[!htb]
\centerline{
\includegraphics[width=.5\linewidth]{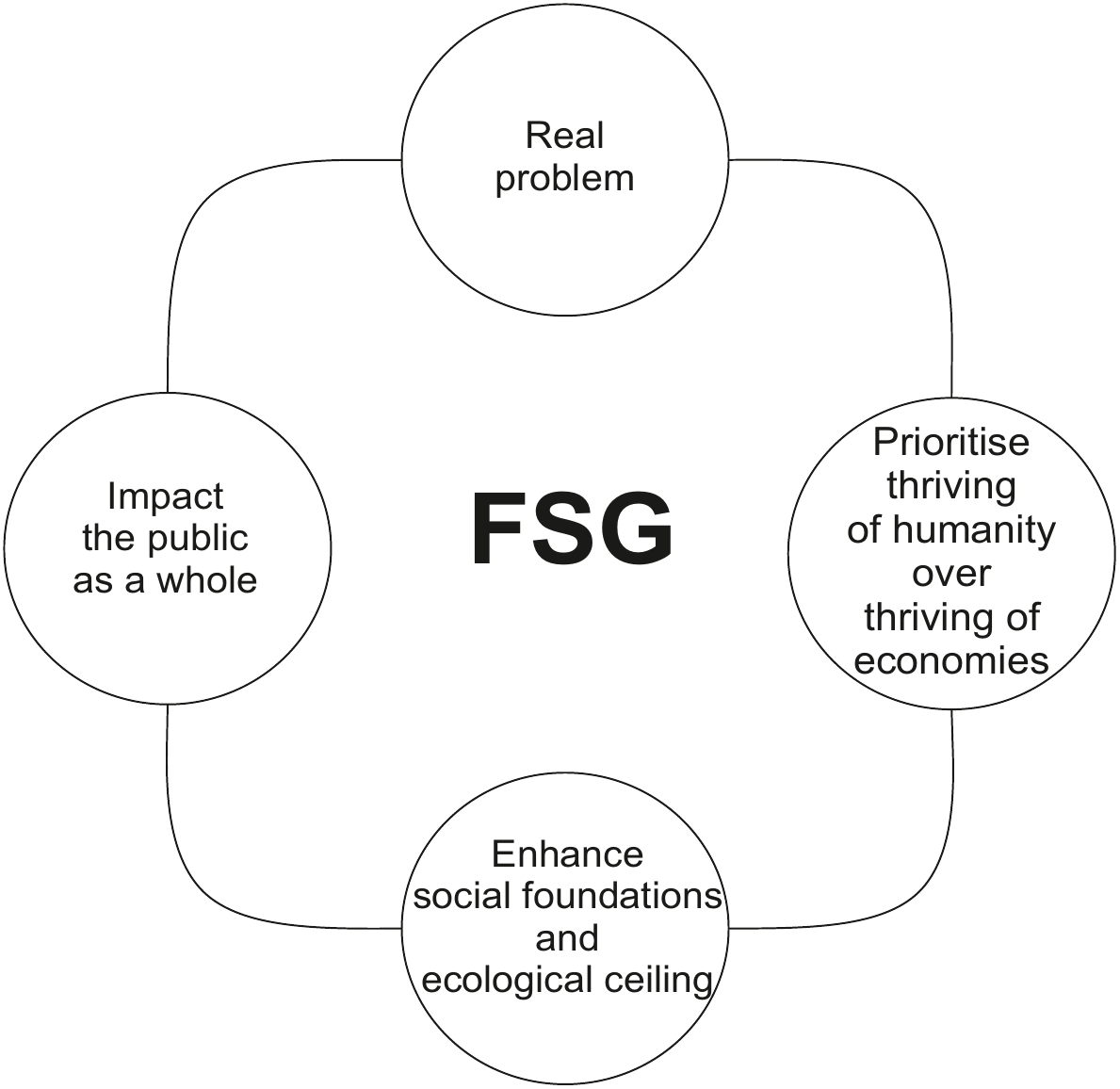}}
\caption{ Attributes of FSG.}
\label{fig_attr}
\end{figure}

These four attributes of FSG can be understood as concerning both the problems driven by thriving humanity and decisions being made in the light of forecasts generated by the forecasting process to enhance social foundation and ecological ceiling, as illustrated in Figure \ref{fig_process}.

Throughout this article we focus on research that substantially relies on forecasting. However, there are other data-driven initiatives related to FSG which might overlap with FSG. Moreover, the forecasting process in FSG might be different compared to other areas of forecasting when it comes to its input, process and output 
In the next subsection we discuss the FSG process and its overlap with other data-driven social good initiatives.

\subsection{Areas related to FSG}\label{sec:other}

\subsubsection{Forecasting process in FSG versus other areas of forecasting}

The unique attributes of FSG discussed in Section \ref{sec:fsg} can lead to various changes throughout the forecasting process including input, process, and output from Fig~\ref{fig_forecasting_process} that is discussed in this subsection.

\textbf{Input}

\begin{itemize}
    \item \textbf{Problem}: As discussed in section 2.1, the forecast problem needs to be real and primarily driven by thriving humanity over economic growth through improving social foundation within the ecological boundaries.
    \item \textbf{Data and Information}: The data and information used in FSG projects can often be more publicly accessible than when there are commercial interests to consider \citep{exchange2020datasets}. However, confidentiality may be required for privacy reasons, especially when the project involves individual-level data. For instance, individual-level data on health, social services or even real-estate prices must be anonymised or made confidential in some way to protect individuals, but data at higher levels of aggregation can often be shared. For example, the aggregated data in healthcare have been shared by the Centre for Disease Prevention and Control \citep{CDCwebsite} in the United States and the National Health Services \citep{NHSwebsite} in the UK. Additionally, we expect to observe lots of missing data, poorly recorded data, the need to combine information from various data sources and data types, and the need for the contextual knowledge of domain applications.
\end{itemize}

\textbf{Process}
\begin{itemize}
    \item \textbf{Software}:  The development of free open-source forecasting software has provided a platform for social good use everywhere. This is because it can be installed and used with no cost for the user while having a huge support from community of users, maintainers and developers. The most widely used open-source forecasting software is the forecast package for R \citep{hyndman2020package}, first released in 2006, and downloaded over 2 million times in 2019. More recently, tidyverts \citep{tidyverts2020} and tidymodels \citep{tidymodels2020} have been introduced for tidy forecasting and modeling. Several other R packages for forecasting are listed on the CRAN Task View for Time Series \citep{TimeSeriespkg}. Another open-source software that has been used to create forecasting tools is Python. Statsmodels library \citep{seabold2010statsmodels} in Python allows for statistical forecasting and scikit-learn library \citep{garreta2013learning} is used more for machine learning. Commercial software such as Oracle, SAP, Simul8, Optima, Tableau, SAS, Forecast Pro and others might also be used in FSG given that they incorporate forecasting modules in their solutions.
    
    \item \textbf{Method}: It is important to note that FSG may or may not involve a novel statistical forecasting methodology. While in some cases societal challenges may lead to innovative research development, the application of existing methods in novel ways is also included in FSG. Moreover, problems in FSG often have small datasets, or in some cases the data is not available at all or the data is incomplete and its quality is unreliable. Therefore, the application of well structured qualitative approaches in such circumstances might be more appropriate. This could also lead to new forecasting methods that concentrate on incomplete and small datasets.
    We should also note that the importance of aligning projects with a real problem in social foundation and ecological ceiling highlights the difference between simply applying exciting forecasting methodologies to a dataset in domain applications and FSG. The latter must have a broader appreciation for the context in which forecasting method would be used in order to provide solutions that can effectively contribute toward achieving the goal.
    In FSG, we are not only interested in a method’s forecast accuracy, but also in its reproducibility, interpretability and transparency. The absence of sufficiently documented methods and computer code underlying the study effectively may undermine their value and becomes a barrier in their use and implementation. \citep{hyndman2010encouraging,boylan2015reproducibility,boylan2016reproducibility,haibe2020importance}. Another part of new methods is developing techniques to estimate model parameters with novel loss functions driven by FSG.
    
    \item \textbf{Estimation}: Ideally the loss function that is used to estimate parameters in the forecast model of FSG should be stated in terms of the decision maker's utility function based on social good metrics rather than statistical measure such as Mean Squared Error and Information Criteria or financial KPIs. An example of a social good loss function in the Emergency Department forecasting would be the use of a loss functions that accounts for patient's waiting time, staff well-being, staff retain, pressure on other health services and costs associated with extra resource
    
    \item \textbf{Evaluation}: The performance of forecasting methods should be evaluated based on metrics of social foundation and ecological ceiling at both local and global levels as disused in Section \ref{sec:scope} rather than measures based on forecast error or financial KPIs.
\end{itemize}

\textbf{Output}

\begin{itemize}
    \item \textbf{Report}: When forecasting is intended to provide social good and to prioritise the public as a whole, the results should be widely reported to maximize the benefit of the forecast. FSG is often going to be of interest to, and hence scrutinized by a wide audience. Thus transparency and trust may emerge as being more important than raw predictive ability. Consider the recent and current discussion of earthquake predictions in Italy \citep{benessia2017earth}, pension dispute in higher education in the UK \citep{wong2018tthe} and the recent COVID19 pandemic \citep{shinde2020forecasting}. In some domains, forecasters can be held liable. Weather forecasts are, for example, widely available on websites, apps and in other media. Modern reporting tools such as Rshiny and Dashboard make it easy to create user-friendly web-based interfaces for reporting forecasts. Example of using Rshiny for FSG include the FluSight Network that shares real-time forecasts of influenza in the US each week, COVID-19 Forecast Hub and modeling COVID-19 \citep{reich2019collaborative,hill2020modeling}. While forecasts specifically designed for the desired application in social good should provide the best information, in some cases forecasts generated for other purposes can be used to provide good information for social good decision making, e.g., climate models can be used for early warning in predicting droughts that can inform humanitarian disaster relief planning \citep{travis2013design,coughlan2015forecast}.
\end{itemize}

\subsubsection{FSG versus other social good initiatives}\label{sec:otherinit}

Forecasting for Social Good is built on previous movements aiming at using technology to positively impact the society. One of the initial movements in that direction is the Tech for Social Good that broadly uses digital technology to tackle societal challenges \citep{Chaudhary2015tech}. Another related area is “Green Supply Chain” that uses a range of technologies and measures to incorporate the ethical and environmental responsibilities into the core culture of contemporary business models \citep{min2012green,zhu2004relationships}. With the increase in data availability in the recent decade and the interest in using the power of data to tackle societal challenges, these initiatives have slowly branched out leading to data-driven initiatives for social good \citep{cuquet2017societal}. Data Science for Social Good (DSSG), Artificial Intelligence for Social Good (AISG), Pro Bono Operations Research (Pro Bono OR) and Statistics for Social Good (SSG) are among the closer related movements to Forecasting for Social Good.

DSSG is defined as “applying data science to improve civic and social outcomes”. The initiative was introduced to help non-profits and government organisations achieve more with their data \citep{moore2019ai}. Several other forms of engagements have since been introduced to derive insights from data in order to help solving social issues. These engagements might be found in the form of fellowships, conferences, competitions, volunteer-based projects, innovation units within large development organizations, and data scientists employed directly by smaller social change organizations. Another similar initiative to DSSG is AISG that focuses on the techniques usually utilised in the Artificial Intelligence field towards social good. DSSG and AISG terms have been used interchangeably in research. Pro Bono OR initiatives aim at connecting OR/analytics professional volunteers with social good causes. Volunteers donate their time and skills to help  nonprofit organizations make better decisions. SSG uses data analysis, statistical and computational techniques to tackle social problems. SSG focuses mainly on problems stemming from economic inequities, like poverty, hunger, human trafficking, and unequal access to education. Table 1 summarises related areas to FSG.

\begin{sidewaystable}
\centering
\footnotesize\setlength{\tabcolsep}{8pt}
  \caption{Related areas to FSG}
  \label{my-label}
\begin{tabular}{lp{.15\textwidth}p{.12\textwidth}p{.2\textwidth}p{.17\textwidth}p{.133\textwidth}}
\toprule
    {Related area} & {Main scope} & {Core techniques} & {Main application domains } & {Some Initiatives} & {Reference}  \\ \midrule
    DSSG  & Governments, nonprofits & Data Science, collect data (and questions), analyze (using visualization and models), communicate & Education, health, criminal justice, sustainability, public safety, workforce development, human services, transportation, economic development, international development, humanitarian, disaster relief operations &
    \href{https://www.dssgfellowship.org/}{Data Science for Social Good}, \href{https://ai-4-all.org/}{AI4ALL}, \href{https://hack4impact.org/}{hack4impact}, \href{https://www.drivendata.org/}{DrivenData}, \href{https://www.datakind.org/}{DataKind}, \href{https://www.unglobalpulse.org/}{United Nations Global Pulse}&
    \cite{chou2014democratizing,catlett2015big,nino2017data,ghani2018data}\\
 \midrule
     AISG  & Governments, nonprofits & Machine learning, Deep learning& Agriculture, education, environmental sustainability, healthcare, combating information manipulation, social care and urban planning, public safety, and transportation & \href{https://www.cais.usc.edu/about/mission-statement/}{USC center for artificial intelligence in society}, \href{https://aiforsocialgood.github.io/2018/}{NeurIPS}, \href{https://aiforsocialgood.github.io/icml2019/}{ICML}, \href{https://aiforsocialgood.github.io/iclr2019/}{ICLR}, \href{https://ai.google/social-good}{Google}, \href{https://fbaiforindia.splashthat.com/}{Facebook}, \href{https://www.ibm.com/watson/advantage-reports/ai-social-good.html}{IBM}, \href{https://www.intel.com/content/www/us/en/artificial-intelligence/ai4socialgood.html}{Intel}, \href{https://www.microsoft.com/en-us/ai/ai-for-good}{Microsoft}, \href{https://www.whitehouse.gov/ai/}{U.S. government}, \href{https://www.baai.ac.cn/en}{Chinese government} &
     \cite{chui2018notes,hager2019artificial,berendt2019ai,shi2020artificial}\\
 \midrule
    Pro Bono OR  & Third sector, Nonprofits & Operations research/ operations management/ management science & Charities, trade associations, credit unions, social enterprises and voluntary organizations & \href{https://connect.informs.org/probonoanalytics/home}{INFORMS}, \href{https://www.theorsociety.com/get-involved/pro-bono-or/}{OR Society} & \cite{mccardle2005or,johnson2007community,midgley2018community}
 \\
     \midrule
    SSG  & Economic inequities, developing nations & Data analysis, statistical and computational techniques & poverty, hunger, human trafficking, and unequal access to education & \href{https://stats-for-good.stanford.edu/}{Statistics for Social Good}, \href{https://swb.wildapricot.org/}{Statistics without borders}& \cite{ashley2010considerations,hwang2019improving} \\ \bottomrule
\end{tabular}
\end{sidewaystable}

DSSG, AISG, Pro Bono OR and SSG are broader terms that may include forecasting as a component. The need for forecasting is driven by uncertainty around the future decisions dealing with societal challenges that need to be made in the light of forecasts.
FSG might differ from these movements in the following ways:

\begin{enumerate}
    \item While DSSG, AISG, Pro Bono and SSG initiatives are defined as domain applications, their scope might be limited to certain organisations or sectors. FSG is not defined based on domain applications, it is inclusive and does not exclude anyone;
    \item FSG is still valid in the absence of data, the area of judgemental forecasting is a valuable tool in the lack of data. However, this is not the case with DSSG, AISG, Pro Bono OR and SSG;
    \item Our focus in FSG is narrowed down from the general data science, artificial intelligence, statistics or operations research, to the use of forecasting for social good improvement;
    \item FSG acts as a compass for the way we do forecasting research and engage with the society at various scales, from an individual to an organisation level;

\end{enumerate}

\section{Research in FSG}\label{sec:quad}

\begin{figure}[!htb]
\centerline{
\includegraphics[width=.6\linewidth]{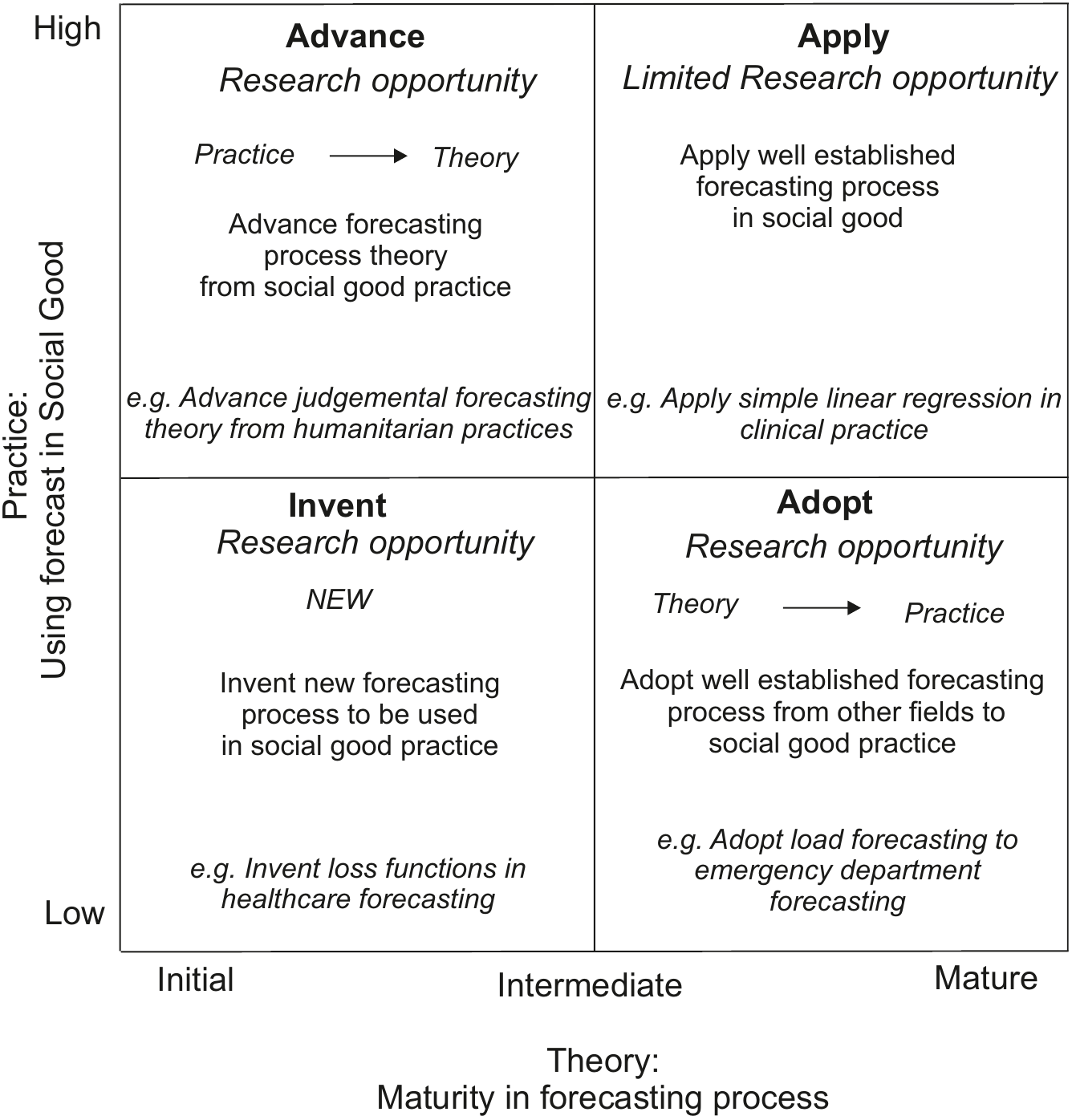}}
\caption{ FSG Research Maturity Framework.}
\label{fig_maturity}
\end{figure}

In this section, we provide a framework that allows the forecasting community, researchers and practitioners to discuss the status of research in FSG and to discover new research opportunities where they can come together to contribute to the area of forecasting for social good.
Figure~\ref{fig_maturity} presents a $2\times 2$ matrix of research maturity \citep{stokes2011pasteur, gregor2013positioning} in FSG based on two dimensions: i) theory: maturity of forecasting process research and ii) practice: use of forecast for social good.

In this framework the forecasting process maturity is defined from initial to mature levels, where:

\begin{enumerate}
    \item \textbf{Initial}: It is characterised by a lower range of topics and methodologies, with a few researchers focusing on the area.
    \item \textbf{Mature}: It is characterised by well-developed forecasting processes that have been studied over time by many researchers resulting in a body of knowledge that contains points of broad agreement.
\end{enumerate}

We consider four areas of development as illustrated in the FSG Research maturity framework in Figure \ref{fig_maturity}. We discuss each quadrant and explore some examples of research opportunities for each one.

\textbf{Apply}

This quadrant is concerned with well established forecasting process research that is regularly used in social good. This implies that users know at least conceptually the forecasting process and how to do it. Therefore, the forecasting process is applied widely across social good as a routine work. Research opportunities and contribution to research might be less obvious but \sout{it is} not impossible. For example, simple linear regression models are widely applied in social good practices such as Medicine, Emergency Department and Emergency Medicine Service to inform policies \citep{boyle2012predicting,kuk2013model}.

\textbf{Adopt}

This quadrant is related to well-defined forecasting processes that are not used widely in social good. We may face situations where the effective forecasting process is not available and used in social good, however it may exist in other areas. Therefore forecasting processes can be adopted, refined or extended for a particular need of social good. It is also possible to adopt a well-defined forecasting process from one application of social good to another. Projects fitting this quadrant provide a great opportunity for research contributions towards applications and possibly knowledge. A large part of research in social good might fall in this quadrant. For instance, successful use of forecasting processes in load demand could be adopted to forecasting emergency department demand as both deal with sub-daily data \citep{rostami2020anticipating}. \citet{van2016demand} employed knowledge available in intermittent demand forecasting theory to forecast humanitarian needs for Medecins Sans Frontieres (MSF-OCA).

\textbf{Advance}

This quadrant focuses on a situation where forecasts -in various forms of estimation- are used in social good, however the forecasting process is not mature. FSG practices can improve the effectiveness of the forecasting process and advance its level of maturity. There are research opportunities here towards contributing to advancing the forecasting process theory. For instance, practices in the area of energy forecasting led to the advance of the theoretical framework of probabilistic load forecasting \citep{hong2016probabilistic}.
In humanitarian and disaster relief operations, experts are using their own experience, expertise and opinion to estimate the humanitarian relief needs and make decisions accordingly. Given the high level of uncertainty such as impact of disaster, its duration, the demand and supply requirement, in the humanitarian and disaster relief forecasting, it is possible that there are methods developed in handling humanitarian and disaster relief operations where multiple perspectives need to be brought together quickly, and these methods may have wider applicability in forecasting problems \citep{unpublishedaltay2020}. Hence, it is likely that FSG practice may lead to improvement and advance research maturity in the judgemental forecasting process. 

\textbf{Invent}

This quadrant concerns innovative forecasting processes that are new to social good. This will contribute to both forecasting process research maturity and the use of forecasts in social good.
For instance, the development of new forecasting methodologies that is directly integrated to the decision making process and its accuracy is evaluated based on social good metrics is an important avenue. An accurate forecasting method evaluated based on statistical measures might not necessarily led to an accurate social good metric. This is because the translation between forecast errors and social good metrics might not be linear. This is a well known issue in forecasting for inventory control \citep{syntetos2009forecasting, kourentzes2020optimising}.
Another example would be identifying appropriate loss functions for social good to estimate the parameters. It is crucial to produce forecasts that are tuned to social good loss functions rather than assuming that the most accurate forecasts based on statistical measures are always best. The social good context has asymmetric and unusual losses that should be taken into account. Forecasting for resource planning is a common task in the health forecasting. A loss function that can balance the over versus under capacity could be used to optimise the forecasting model parameters.
Finally, the limited capacity to record data in developing countries and the data quality issues related to that, especially when it is coupled with humanitarian crises is very common. In this context, other similar humanitarian disasters may have data that could be applied to a new disaster/event. Therefore, developing new forecasting processes that specifically focus on small and messy datasets in social good is important.

We should note that the FSG research maturity framework is not prescriptive. It can serve as a tool to help researchers and practitioners map their research to social good practices. This will help them to prioritise their research agenda, identify areas where they can contribute to social good and create opportunities to advance FSG knowledge and close the gap between theory and practice in FSG.

\section{Conclusion}\label{sec:concl}

Forecasting is an integral part of organisational decision making, but its linkage to non-economic/financial utility has been limited. Better integration of forecasting with environmental and social KPIs is both feasible and desirable, and relevant practices have been receiving increasing attention as a means to safeguarding and generating social good. With the support of the International Institute of Forecasters (IIF), forecasting for social good (FSG) has recently been introduced as a self contained area of scholarship, enabling focused academic research and facilitating a constructive exchange of ideas between academia and the private and public sector \citep{fsg2018cardiff,fsg2021bordeaux}. 

In this paper, we have attempted to further formalise FSG in order to increase awareness and interest of academics and practitioners on its potential impact; encourage interested academics and practitioners to engage in this important agenda; and inspire the development of new forecasting methodologies tailored for social good applications.

We find the Doughnut theory accommodating, towards reaching a helpful definition of FSG: it is concerned with real social problems both in terms of application and performance measurement, and emphasises society as a whole. Different from other data science, statistics, and  operations research initiatives that emphasise social good, FSG is not restricted to particular organisational contexts or sectors, and capitalises on the fundamental advancements that have been made in the area of judgmental forecasting, to dissociate substantive contributions from the availability of (quantitative/hard) data. Mapping the maturity of research in various areas of forecasting against FSG practice allows us to identify opportunities for bridging the gap between the theory and practice of FSG. When practice lags behind theory, there is an opportunity to adopt already existing theory to advance practical applications. When theory lags behind practice, there is a need to advance forecasting research, building on the insights and lessons learned from practical applications. The forecasting community is called to invent new approaches in areas where neither sufficient knowledge nor empirical evidence have been accumulated.

The FSG guidelines we present in this paper are not intended to be definitive, and we recognise that relevant work may indeed fall outside our working framework. The intention of FSG is to motivate engagement with important issues facing our world and society and allow best (forecasting) practices to emerge. That is, we hope a definition of FSG and its introduction as a self-contained area of inquiry will lead to increased appreciation of forecasting as an enabler of greater social good. Qualifying what constitutes FSG should permit academics and practitioners to appreciate the opportunity cost of not engaging with its scalable agenda.

There are a number of ongoing initiatives in this area \citep{dssg2020,usc2020}, including dedicated workshops \citep{fsg2018cardiff,fsg2021bordeaux}, International Journal of Forecasting special sections \citep{ijsfsg2018,ijsfsg2020}, invited sessions in the International Symposium on Forecasting \citep{rostami2019isf}, and some longer term work led by the first author of this paper on Democratising Forecasting \citep{democratisingforecasting2020}, a project the goal of which is to provide forecasting training to individuals in developing countries around the world. Just like FSG, this is born from a recognition of the benefits that forecasting tools can bring to advancing social justice goals. However, it goes one step further in not only making a connection between forecasting and its social utility, but emphasising direct capacity building and improving forecasting expertise in deprived economies.
We hope our paper will motivate and inspire forecasting experts to put their knowledge to a good cause and we look forward to relevant developments in the years to come.

\clearpage

\bibliographystyle{agsm}
\bibliography{fsg}

\end{document}